# Study of Glass and Bakelite properties as electrodes in RPCs

Manisha*, V. Bhatnagar, J.S. Shahi

Department of Physics, Panjab University
Chandigarh-160014, India

## *Abstract*

India-based Neutrino Observatory (INO) collaboration is planning to build a magnetized Iron-CALorimeter detector (ICAL) for the study of atmospheric neutrinos. ICAL detector will be a stack of 151 layers of magnetized iron plates interleaved with Resistive Plate Chambers (RPCs) as active detector elements with a total mass of 50 kton. Resistive Plate Chambers are gaseous detectors made up of two parallel electrodes of high bulk resistivity like float glass and bakelite. These detectors are extensively used in several high energy physics experiments since 1980s because of high count rate, excellent time as well as spatial resolutions, simple to fabricate and operate. Due to detector aging issue, it is necessary to characterize electrode material so as to select appropriate electrode material before fabricating the detector. In the present studies, we measured bulk resistivity and surface current of glass as well as bakelite. Bulk resistivity of bakelite is ~ 100 times less than that of glass and surface current of bakelite is higher than that of glass. Also glass does not need any kind of surface treatment to achieve better surface uniformity. Therefore, glass electrodes are preferred over bakelite electrodes in most of the cases. Locally manufactured Asahi glass of ~2 mm thickness and bakelite sheets were tested during the studies as reported in this paper before the various stages of detector fabrication.

---

* Corresponding author : manisha@pu.ac.in

# 1. Introduction

India-based Neutrino Observatory (INO) [1] is a proposed underground multi experiment facility aimed to study atmospheric neutrino physics extensively i.e. to reconfirm occurrence of oscillation in atmospheric neutrinos already confirmed by Super K experiment, precision measurements of oscillation parameters related to atmospheric neutrinos [2,3]. Detector required for such an experiment is supposed to have large target mass to achieve statistically significant number of neutrino interactions for reconfirmation of atmospheric neutrino oscillations [4], also good energy resolution and identification of electric charge of muons so as to distinguish between neutrinos and anti-neutrinos. ICAL not only satisfies the criteria required for such a detector in such an experiment, also compared to other detectors ICAL have a large range in sensitivity to path length (L) and neutrino energy (E) as they appear in the probability relation for oscillation [5]. ICAL will have modular structure consisting of 3 modules, one ICAL module will be a stack of iron absorbers interleaved with an air gap of 4 cm housing RPCs [6,7] as active elements. Therefore, it is necessary to choose an appropriate electrode material for RPC fabrication [8]. We perform bulk resistivity as well as surface current studies of IEL bakelite, Formica bakelite, Hylam bakelite, Italian bakelie and Asahi glass of ~ 2 mm thickness procured from local market. Based on the studies performed for bulk resistivity and surface current of Asahi glass sample and various bakelite samples (as mentioned above), we choose Asahi glass as electrode for RPC fabrication.

# 2. Bulk resistivity and surface current measurements

Bulk resistivity and surface current measurements of various bakelite samples and Asahi glass sample of ~ 2 mm thickness were performed using standard Two Probe setup. Two Probe setup (shown in figure 1) consists of Digital Picoammeter, High Voltage Power Supply and Two Probes arrangement. Two Probes arrangement consists of two spring loaded contact probes moving in pipes and insulated by Teflon washers. One of the probes would be in contact with upper surface of the sample under test (resting on base plate); other probe would be in contact with lower surface of the sample through Al foil as shown in figure 2. This probes arrangement is mounted in a suitable stand which also holds the sample plate. Teflon coated leads are provided to connect this probes arrangement with High Voltage Power Supply and Picoammeter.

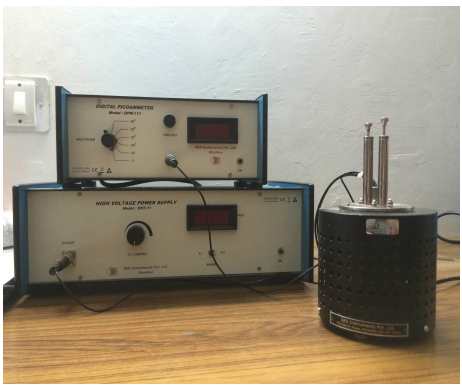
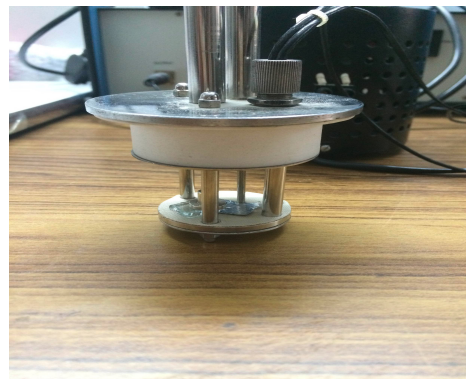

Figure 1 : Two Probe setup.　　　　　Figure 2 : Two Probes arrangement.

Measured bulk resistivity is ~ $10^{10}$ Ω-m (as shown in figure 4), surface current is ~ few nA (as shown in figure 3) for glass as well as various bakelite samples.

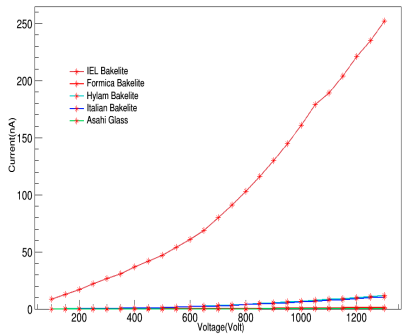 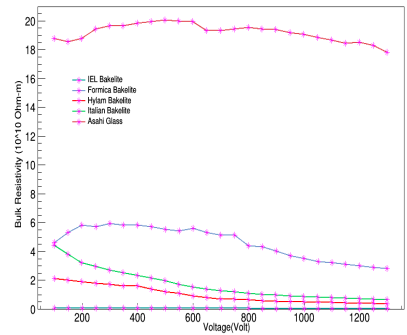

Figure 3 : Surface current vs. voltage characteristics for glass, bakelite samples.

Figure 4 : Bulk resistivity vs. voltage characteristics for glass, bakelite samples.

## 3. Glass RPC fabrication

Procuring Asahi glass plates of ~ 2 mm thickness, we fabricate glass RPC. To fabricate glass RPC, firstly glass plates were cleaned with alcohol and distilled water as shown in figure 5. Inserting gas nozzles on four corners, polycarbonate edge spacers were glued on one of the glass sheet using 3M Scotch-Weld Epoxy adhesive glue as shown in figure 6. Once these nozzles and edge spacers were strengthened, then 16 button spacers of polycarbonate having thickness ~ 2 mm were glued using mylar template as shown in figure 7. One of glass RPC gap is shown in figure 8.

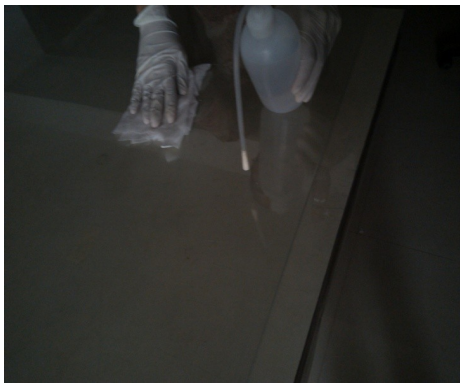 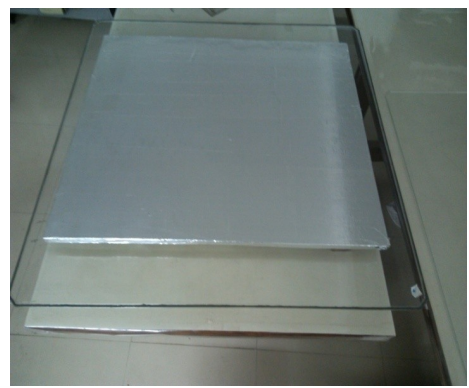

Figure 5 : Cleaning glass with alcohol, distilled water.

Figure 6 : Glued edge spacers with glass sheet.

## 4. Physical tests

Before proceeding further, it is necessary to perform leak test and spacer test of RPC gas gap. These tests also specify mechanical stability of RPC. To perform these tests, we use Ar gas. Ar gas was flown through RPC under controlled pressure of 5-6 mbar. Once Ar gas flow is on, for first 10-15 minutes, it will take time to stabilize the gas flow. For next 15-20 minutes, the rate of change of pressure (*dP/dt*) with time can be determined. If rate of change of pressure is ~ $10^{-5}$ mbar/second, then RPC gas gap under test is fine. If rate of change of pressure is more than it, then that RPC gas gap is leaky. Once leak rate is determined, then start gently pressing all the glued button spacers one by one at the marked positions. As button spacer is pressurized with hand, it will immediately show a peak. If all the peaks are of the same level, it means all the button spacers are properly glued. If any of the peaks overshoots, it means corresponding spacer is popped up. A schematic diagram of leak & spacer test setup to perform aforesaid tests is shown in figure 9.

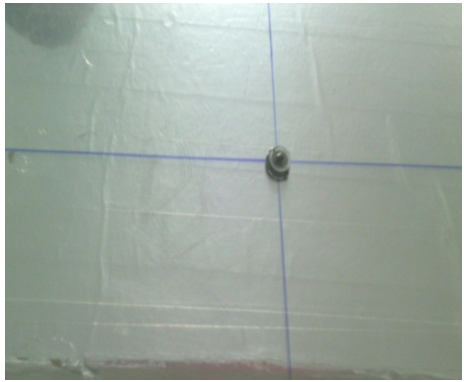

Figure 7 : Mylar template to glue button spacers.

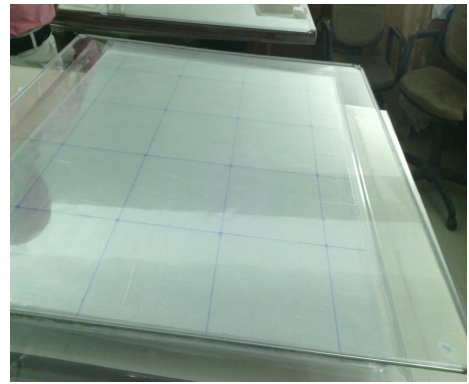

Figure 8 : Glass RPC gap.

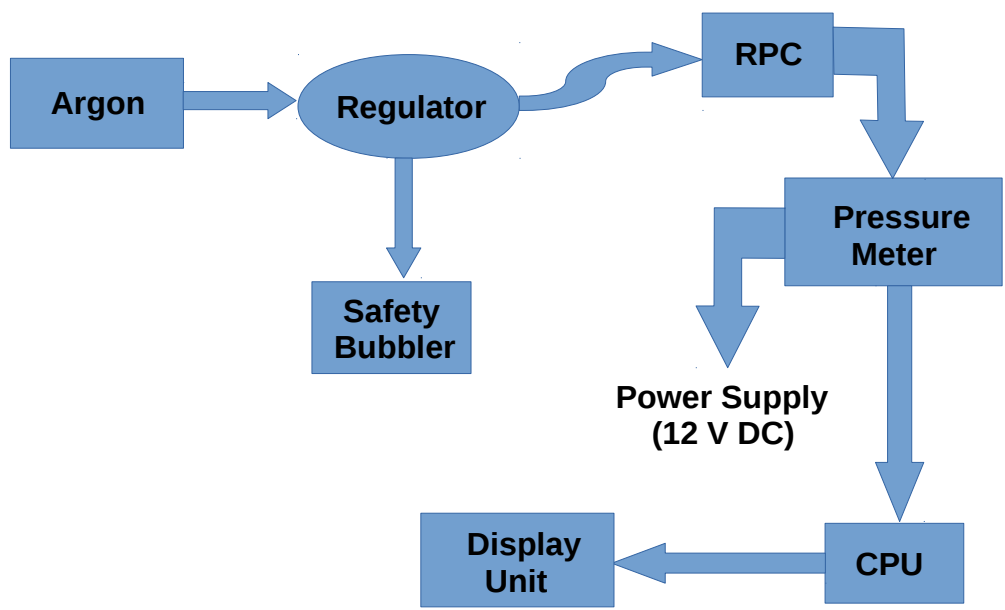

Figure 9 : Schematic diagram of leak and spacer test setup.

If RPC gap qualifies both spacer & leak test, then it is coated with graphite paint of well defined composition to make highly insulating glass surface conductive.

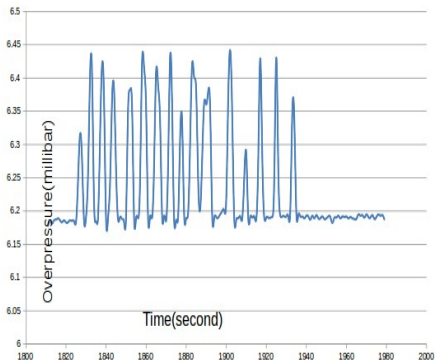
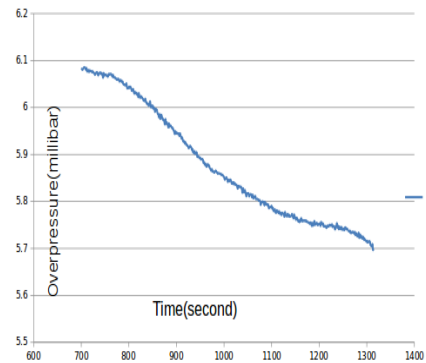

Figure 10 : Spacer test plot of glass RPC.     Figure 11 : Leak test plot of glass RPC.

## 5. Surface resistivity measurements

Surface resistivity measurements of graphite coated glass RPC gap are carried out using multimeter, zig made of copper bars (~ 16 mm thick) and stainless steel scale (used for alignment purpose) as shown in figure 12. Surface resistivity measurements were carried out in both horizontal as well vertical mode of measurements. Two modes of measurements vary only in the way of moving the zig. Resistivity plots obtained after resistivity measurements in horizontal and vertical mode are shown in figure 13, figure 14 respectively.

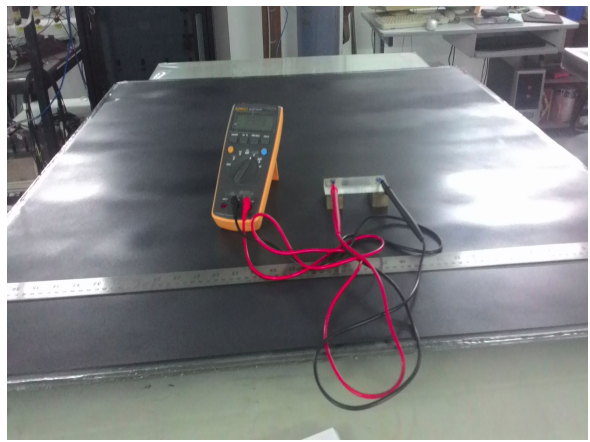

Figure 12 : Surface resistivity measurements of graphite coated glass RPC gap using zig.

Uniformity in graphite coating is one of the important parameter on which quality of induced signal depends. Therefore we calculate % variation in graphite coating by choosing alternate blocks of RPC gas gap surface to determine the extent of variation throughout the surface. If variation is less than or equal to 10 %, then it is accepted. If % variation is more than 10 % then graphite coating is cleaned using thinner and gap is coated again. This coating and extent of variation measurements, both are repeated again and again until or unless variation in surface resistivity measurements is in the accepted range. Calculated values of % variation for both the aforesaid measurements are shown in Table 1.

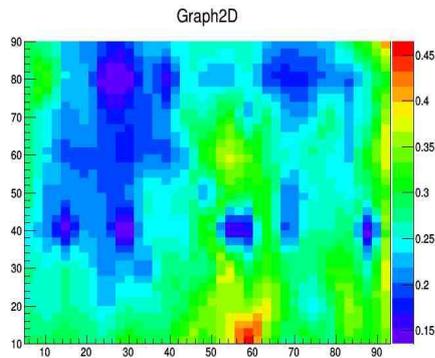

Figure 13 : Horizontal resistivity measurements.

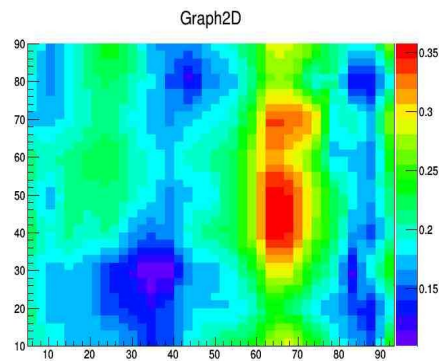

Figure 14 : Vertical resistivity measurements.

Table 1 : % Variation in measured surface resistivity values.

| Mode of measurements | No. of resistivity values used | Mean resistivity | Standard deviation | % Variation |
| --- | --- | --- | --- | --- |
| Horizontal Measurements | 50 | 0.278 | 0.0584 | 5.84 % |
| Vertical Measurements | 50 | 0.202 | 0.0475 | 4.75 % |

## 6. Results and conclusions

The functioning of the RPC, as a gaseous detector based on ionization, is highly affected by the choice of electrodes. Glass and bakelite, both being used as electrodes having high resistivity (~ $10^9$-$10^{11}$Ω-m) play major role in localizing the discharge only up to a limited area in the RPC. Such highly resistive electrodes are made conductive, for inducing charge on the signal pick-up panel, with the help of graphite coating bringing surface resistivity in the range of 0.1-1 MΩ. The presented studies verify that glass and bakelite have the required properties to fulfill the need of RPC's electrodes. There are constructional and operational differences between glass and bakelite based RPC, which are not part of this presented work.

## Acknowledgements

We would like to thank Department of Science and Technology (DST)/Department of Atomic Energy (DAE) to provide financial support for R&D work. We also thank staff of EHEP group and Department of Physics, Panjab University, Chandigarh.